\documentclass[aps,preprint,preprintnumbers,nofootinbib,showpacs]{revtex4}

\def\beq{\begin{equation}}
\def\eeq{\end{equation}}
\def\beqa{\begin{eqnarray}}
\def\eeqa{\end{eqnarray}}

\def\lsim{\mathrel{\raise.3ex\hbox{$<$\kern-.75em\lower1ex\hbox{$\sim$}}} }
\def\gsim{\mathrel{\raise.3ex\hbox{$>$\kern-.75em\lower1ex\hbox{$\sim$}}} }

\begin{document}
\preprint{\tighten{\vbox{\hbox{NCU-HEP-k009}
\hbox{rev. Jul 2003}
}}}
\vspace*{1in}

\title{A Completed Chiral Fermionic Sector Model with Little Higgs
\vspace*{.3in} }
\author{\bf Otto C. W. Kong \vspace*{.2in}}
\email{otto@phy.ncu.edu.tw}
\affiliation
{ Department of Physics, National Central University, Chung-li, Taiwan 32054\\
\& \ \ \ \ Institute of Physics, Academia Sinica, Taipei, Taiwan 11529 
\vspace*{.5in}}

\begin{abstract}
The Standard Model has some intrinsic beauty in the sector of fermions and gauge bosons. Its
scalar sector, though minimal, is however haunted by the hierarchy problem.  The fermionic
spectrum also have two major problems, the flavor problem with its fundamental notion about
why there are three families, and the phenomenological limitation of massless neutrinos. 
We present here a completed chiral fermionic sector model, based on a little Higgs model, that
has the plausible potential of addressing all these problems of the SM at an accessible energy
scale, and comment briefly on its phenomenology. The focus here is not on the little Higgs part, 
but rather on the electroweak quarks and leptons from the model, which of course from an
important part of the full model. 
\end{abstract}
\maketitle

Particle theorists have been working on extending the Standard Model (SM) for some
thirty years. A major theme in such model-building works is to extend the 
$SU(3)_C\times SU(2)_L\times U(1)_Y$ gauge symmetry, as exemplified by the
classic $SU(5)$ grand unification model\cite{su5}. The attempts to go beyond the SM are 
motivated by many limitations of the SM itself, as much as by our desire to see some
other layer of structure in nature. For that matter, the grand desert spanning the next fourteen 
orders of magnitude in energy/length scale as suggested by the (supersymmetric) gauge
unification idea certainly sounds a bit boring to some of us. Does nature has more excitement 
to offer at the next energy/length scale? Supersymmetry (SUSY) is a beautiful idea, but its low
energy implementation does leave much to be desire. However, the supersymmetric SM
stays popular over the decades for some good reasons. In our opinion, the major part of it
is the fact that it offers a solution to the hierarchy problem (of stabilizing the Higgs masses)
without quite compromising the perfect beauty of the (one-family) chiral fermionic spectrum
of the SM, while maintaining its phenomenological viability in front of all the precision
electroweak data. All those are done by incorporating one basic symmetry --- supersymmetry
into the SM. To contemplate a really competitive alternative, one needs to take this to the heart.

The first central ingredient of the SM, or its supersymmetric extension, is its gauge symmetry.
One perspective of extending that gauge symmetry leads to the ``exceptional"  embedding
sequence of
\[
SU(3)_C\times SU(2)_L\times U(1)_Y \; \subset \; SU(5) \; \subset \; SO(10)
\; \subset \; E(6) \; \subset \; E(7)  \; \subset \; E(8) \;.
\]
From the perspective of the fermionic spectrum, the embedding is vertical, {\it i.e.} a
one-family unification. The highly nontrivial chiral gauge anomaly cancellation among
the different fermionic multiplets of the quarks and leptons in the SM is sort of trivialized
in the embedding at the $SO(10)$ level, while new fermionic states also started to be required.
The most intriguing aspect of the spectrum, the fact that there are three families, remains 
unexplained. Adding SUSY only make the flavor problem more complicated from the 
phenomenological point of view. There are also in the literature many horizontal
(or family) symmetry models\cite{hs}, in which extra symmetries are added to describe flavor
physics. However, it is fair to say that only such kind of models with a nonabelian gauge 
horizontal symmetry address, to some extent, why there are three families. This is again through
the issue of  nontrivial chiral gauge anomaly cancellation\cite{unc}.

Nontrivial three-family  [$SU(5)$ based] unification models also has an early history\cite{su7}. 
Attempts in model-building with an extended gauge symmetry that incorporates a nontrivial
family structure are however less popular. Such models, if not unification based, should be more
interesting. They could provide much more accessible phenomenology while explaining the
flavor structure. And such (flavor) models offer the potential of tackling the hierarchy problem and 
other limitations of the SM at the same time  --- the new model we present below may be the first 
example with some success along the line. We would like to name two particular examples of
the kind of models, an $SU(3)_C\times SU(3)_L\times U(1)_X$ ({331}) model\cite{331} without
family changing gauge boson, and an  $SU(4)_A\times SU(3)_C\times SU(2)_L\times U(1)_X$
model and generalizations\cite{unc67} with family changing gauge bosons. 

One other limitation of the SM is its massless neutrinos. To explain the neutrino oscillation data, 
we need beyond SM properties of neutrinos --- either couplings with or without extra (singlet) 
neutrino states. While it can be argue that the generic supersymmetric SM, without the {\it ad hoc}
R parity imposed, naturally solve the problem\cite{as89}, most alternatives requires having extra
singlet  (or called right-handed) neutrino states at some scale. 
 
We now start the discussion of the construction of a new model that has the potential to address 
the hierarchy problem, flavor problem, and neutrino mass problem, altogether at an accessible
energy scale. The desirability of such a model is well illustrated by the above discussions. The
starting point of our construction is a simple, though less than perfect, little Higgs model from
Ref.\cite{KS}. 

The little Higgs idea is an interesting alternative solution to the hierarchy problem\cite{S}. The 
SM Higgs boson is here to be identified as pseudo-Nambu Goldstone bosons (PNGB) of some 
global symmetry. Two separate global symmetries, each to be broken by a Higgs vacuum 
expectation value (VEV), are to be arranged such that a 1-loop (SM) Higgs mass diagram is 
protected by the (residue) symmetries to be free from quadratic divergence. The idea 
was motivated by dimensional deconstruction\cite{dd}, though the 
mechanism may not necessarily follows from the strong interaction dynamics behind as 
suggested\cite{L}. Recently, simple group theoretical constructions of little Higgs models are 
attempted\cite{KS,S}. little Higgs construction looks a bit like a heavy machinery to fix only the 
hierarchy problem, and would remain a toy model without a completed description of the SM 
fermionic embedding. The necessarily extended gauge symmetry, to provide quadratic divergent 
Higgs mass contribution cancellation to the electroweak gauge bosons, means that the embedding
into a consistent model is highly nontrivial, due to gauge anomalies. Building such a model
embedding looks similar to the model-building works for the flavor problem discussed above.
This is the major focus of the present letter. The consistent fermionic spectrum is also the basis
for further studies of what implications the little Higgs structure really have for flavor physics.

We take the simplest model with a $SU(3)_L\times U(1)_X$ gauge symmetry in Ref.\cite{KS}.
The $t$-$b$ quark doublet of $SU(2)_L\times U(1)_Y$ is to be embedded into a 
$SU(3)_L$ triplet as follows
\beq \label{tbT}
3_{\!\scriptscriptstyle L} = 
\left[ \begin{array} {c}
t^a \\ b^a \\ T^a
\end{array} \right]
\eeq
with $X$-charge being $\frac{1}{3}$.  The third state is another top-like quark $T$,
with the usual electric charge of ${\mathcal Q}=\frac{2}{3}$; and $a$ represents the 
$SU(3)_C$ index. In fact, we have here
\beq \label{Q}
{\mathcal Q} = \frac{1}{2} \, \lambda_{\!\scriptscriptstyle L}^3 
 - \frac{1}{2\sqrt{3}} \, \lambda_{\!\scriptscriptstyle L}^8 + X \;;
\eeq
or  $Y= X - \frac{1}{2\sqrt{3}} \, \lambda_{\!\scriptscriptstyle L}^8 $ (${\mathcal Q}=T_3+Y$).
The vector like QCD spectrum is to be recovered by having the Dirac partners in 
$SU(3)_L\times U(1)_X$ singlets as
\beq
1_{\!\scriptscriptstyle L} =  \bar{b}_a\;,
\quad  \bar{t'}_a\;,
\quad \bar{T'}_a\;.
\eeq
with ${\mathcal Q}=X=\frac{1}{3}\;\mbox{and}\; -\frac{2}{3}$ respectively.
So far, this is what had been suggested in Ref.\cite{KS}. The extra top quark $T$ is 
exactly what is needed to cancel the 1-loop quadratic divergence in the Higgs mass(es) as 
required by the little Higgs idea. The electroweak Higgs doublets are embedded into 
$SU(3)_L\times U(1)_X$ anti-triplets  ($\Phi_i$'s) of $X=\frac{1}{3}$. The $\Phi_i$'s bear
VEVs that break the gauge symmetry to that of the SM at scale $f$ of about 1 TeV.
One expect the Yukawa couplings
\beqa 
{\mathcal L}_{top} &=& y_{\!\scriptscriptstyle 1}\,\bar{t'}_a\,
\Phi_{\!\scriptscriptstyle 1} \, Q^a   +    y_{\!\scriptscriptstyle 2}\,\bar{T'}_a\,
\Phi_{\!\scriptscriptstyle 2} \, Q^a
\nonumber \\  &=&
f\,(y_{\!\scriptscriptstyle 1}\,\bar{t'}+  y_{\!\scriptscriptstyle 2}\,\bar{T'})\, T
+ \frac{i}{\sqrt{2}} \, (y_{\!\scriptscriptstyle 1}\,\bar{t}'-  y_{\!\scriptscriptstyle 2}\,\bar{T}')\,h
\left( \begin{array}{c} t \\ b \end{array} \right) + \cdots
\nonumber \\ &=&
m_{\!\scriptscriptstyle T}\, \bar{T} T - i y_{\!\scriptscriptstyle t} \, \bar{t}\, h
\left( \begin{array}{c} t \\ b \end{array} \right) + \cdots
\label{tykw}
\eeqa
Here $Q^a$ denotes the triplet of Eq.(\ref{tbT}); and we suppress the color indices
after the first line. Both $y_{\!\scriptscriptstyle 1}$ and $y_{\!\scriptscriptstyle 2}$ 
are expected to be of order one to produced the phenomenological top mass from
electroweak symmetry breaking. The part of the model discussed in this paragraph is all
taken from Ref.\cite{KS}. We have no much to add, in this short letter, to the little Higgs story
but rather prefer to focus on the fermionic part --- the full set of SM quarks and leptons.
Without further dwelling on the little Higgs part of the model, we refer the readers to 
Ref.\cite{KS} for the details of how the quadratic divergence cancellation works here. 
We would like to bring the readers attention to a potential limitation of the current model
as a little Higgs model though. This is the difficulty with getting a good Higgs quartic
coupling. In fact, the latter motivated the authors of Ref.\cite{KS} go to the construction 
of a similar but more preferable  $SU(4)_L\times U(1)_X$ model.  We will report of
constructions of consistent $SU(4)_L\times U(1)_X$ fermion spectra in a coming 
publication\cite{010}.

A question of paramount importance not handled in Ref.\cite{KS}  is the complete
fermionic spectrum under  $SU(3)_L\times U(1)_X$.\footnote{In fact, Ref.\cite{KS} has a 
bit of discussion on the other SM fermions, which we find unsatisfactory from the current 
model-building point of view.}           We emphasize again that the issue 
is nontrivial, due to the chiral gauge anomaly cancellation required. For example, it is 
quite obvious that a similar quark embedding for the lighter two families does not work. 
While all representation of $SU(2)$ are real, the complex nature of the $SU(3)_L$ 
fundamental representation process nontrivial anomaly. The nine 
$3_{\!\scriptscriptstyle L}$ representations has anomaly added up that required 
maybe the same number of $\bar{3}_{\!\scriptscriptstyle L}$ fermions to cancel it.
Adding the large number of fermionic states to the spectrum is highly undesirable. 
Moreover, one would still has to add more states to take care of the SM leptonic
sector. It is not simpler about adding enough representations under the full gauge
group $SU(3)_C\times SU(3)_L\times U(1)_X$ to incorporate all the SM chiral
fermions and taking care of the $SU(3)_L$ chiral anomaly either. The full fermionic
spectrum better be vectorlike at the QCD and QED level or risk predicting the 
existence of extra massless fermions. And finally, all chiral anomalies, including also
the $SU(3)_C$ and $U(1)_X$ parts and all the mixed anomalies, have to be canceled.
If a judicially chosen fermionic spectrum satisfying the above cannot be found,
the little Higgs model remains a toy model for the scalar sector, incapable of
being a consistent particle physics model extending the SM.

However, the family universal structure mentioned above is not necessary, nor
desirable. The idea of treating the third family different from the lighter two has been
a favorable tool in the kit of family symmetry model-builders, as mentioned in the 
introduction above. With the complex $3_{\!\scriptscriptstyle L}$, one may also 
used  a $\bar{3}_{\!\scriptscriptstyle L}$ to house a SM doublet. The bonus is
that we can have nontrivial anomaly cancellation among the three families, hence
bring another major goal of model-builders into the little Higgs game. 

Here, we present in this letter a completed chiral fermionic sector model having
exactly the above discussed features. We have the quark doublets of the other
two families embedded as follow:
\beq
\bar{3}_{\!\scriptscriptstyle L} = 
\left[ \begin{array} {c}
d^a \\ u^a \\ D^a
\end{array} \right] \;,
\quad
\left[ \begin{array} {c}
s^a \\ c^a \\ S^a
\end{array} \right] \;.
\eeq
$X=0$ gives the right $U(1)$ charges. The new quarks $D$ and $S$ are also 
down-type quark (with  ${\mathcal Q}=-\frac{1}{3}$). Each of the quark states has 
a matching Dirac partner to keep QCD vectorlike, {\it i.e.} we have the singlets
\beq
1_{\!\scriptscriptstyle L} =
  \bar{u}_a\,,\;
 \bar{d}'_a\,,\;
 \bar{D}'_a\,,\;
  \bar{c}_a\,,\;
 \bar{s}'_a\,,\;
 \bar{S}'_a\,;
\eeq
each with their $X$-charge given exactly by the electric charge ${\mathcal Q}$
[{\it cf.} Eq.(\ref{Q})]. We have then, within the quark sector, three 
$\bar{3}_{\!\scriptscriptstyle L}$ in excess. The latter is to be canceled by 
three (colorless) ${3}_{\!\scriptscriptstyle L}$ representations housing the
leptonic doublets in a family universal pattern. Explicitly, we have
\beq \label{L}
{3}_{\!\scriptscriptstyle L} = 
\left[ \begin{array} {c}
\nu_e \\ e^- \\ N_e
\end{array} \right] \;,\quad
\left[ \begin{array} {c}
\nu_\mu \\ \mu^- \\ N_\mu
\end{array} \right] \;,\quad
\left[ \begin{array} {c}
\nu_\tau \\ \tau^- \\ N_\tau
\end{array} \right] \;,
\eeq
with $X=-\frac{1}{3}$. The full spectrum is completed with the three leptonic
singlets as 
\beq
1_{\!\scriptscriptstyle L} = e^+\;,\quad
\mu^+\;,\quad
\tau^+\;,
\eeq
with $X=1$. 

One can easily check that all the potentially dangerous triangle anomalies, 
$(3_{\!\scriptscriptstyle C} )^3$, $(3_{\!\scriptscriptstyle C} )^2 X$,
$(3_{\!\scriptscriptstyle L} )^3$, $(3_{\!\scriptscriptstyle L} )^2 X$,
$X$-trace, and $X^3$ --- notation self explanatory, do cancel. We have illustrated
the cancellation of the $(3_{\!\scriptscriptstyle L} )^3$ anomaly in the
construction above. The $(3_{\!\scriptscriptstyle C} )^3$ anomaly is absent for 
we do require the QCD spectrum to be vectorlike. The nontrivial 
$(3_{\!\scriptscriptstyle L} )^2 X$ anomaly contribution from the $Q^a$
triplets is canceled by that from the three families of leptonic triplets
 [{\it cf.} Eqs.(\ref{tbT}) and (\ref{L})]. The cancellation of the remaining anomalies 
are quite nontrivial and take some algebra. The basic feature of cancellation among 
the three families persists. In particular the $(3_{\!\scriptscriptstyle L} )^3$, 
$(3_{\!\scriptscriptstyle L} )^2 X$, and $X^3$ anomalies for each family is nonvanishing. 
The spectrum and anomaly cancellation structure share quite a bit of similarity with
the {331} model\cite{331}, which is a major inspiration for our model construction.
Apart from the little Higgs perspective incorporated in our new {331} model, the 
are other major difference between the two fermionic model spectra. The present
fermionic spectrum is, arguably, phenomenologically more interesting --- an aspect
that we will then turn our discussion to. 

Before going into the different phenomenological predictions of our new  {331} model,
however, we will discuss briefly the similarity in the group theoretical structure it shares
with the original  {331} model\cite{331}. This may shed some light on model-building.
To put it in an oversimplifying statement, one can say that our model is nothing more than
the original {331} model-building idea with a twist --- a different enlargement of the quark 
representations dictated by our interest in accommodating the little Higgs. Taking the 
$Q^a$ triplet with the {331} construction scheme, the rest more or less follows. The
nontrivial anomaly cancellation among the {\it three} families is like intrinsic to the
basic strategy of the $SU(2)_L\times U(1)_Y$ into $SU(3)_L\times U(1)_X$ embedding
with new quarks. It works, largely because the $3$ in $SU(3)_C$ is the same as the
number of SM families. The fact that such construction works, in
either case, is quite amazing.\footnote{See, however, detailed analysis of the guage
anomaly structure of the kind of SM embeddings in our forthcoming publication\cite{010}.}

A major phenomenological different between our new model and the original {331} model
is the fermionic content. The old {331} model has extra quarks of electric charges 
$\frac{5}{3}$ and $-\frac{4}{3}$ as the only fermions beyond the three-family SM spectrum.
In our model, there is the same number of extra quarks. They are, however, just some
duplications of the existing $t$, and $d$ and $s$ quarks --- or rather two more down-sector quarks,
without exotic electric charges. Recall that the extra top quark $T$ is demanded by the little
Higgs mechanism\cite{KS}. There has been many discussions in the literature about extra
down-sector quarks from both theoretical\cite{down-t} point of view, as well as experimental
in which one tries to explain the issues related to the $b$ quark $Z$-width\cite{down-e}.
Hence, the different quark content of the new model actually looks very desirable. In the new 
model, we have also extra leptonic states --- three new neutral fermions $N_e$, $N_\mu$,
and $N_\tau$, which are essentially singlet neutrinos. This, together with extra interactions
of the SM neutrinos, may provide the base for  interesting beyond SM properties of neutrinos.
The gauge boson sector of course plays an important role in the little Higgs mechanism, and 
is also well discussed in that aspect in Ref.\cite{KS}. While  the original {331} model has 
doubly charged gauge bosons that may provide interesting experimental signature\cite{331},
the present model has no extra gauge bosons of exotic charges. The five extra gauge bosons
are rather a pair of $W'$ and three extra $Z'$'s. 

Next we look into the possible couplings of the scalar $\Phi_i$ multiplets to the fermions.
Such couplings are responsible for the SM Yukawa couplings, and the basic properties of the
extra singlet quarks and neutrinos. Besides the top, the bottom quark has to get its Yukawa
coupling from the $\bar{b}\,\Phi_i^\dag \Phi_j^\dag \, Q^a$ term, hence the desired 
suppression in its mass (after electroweak symmetry breaking). For the first two families
the $1_{\!\scriptscriptstyle L} \, \Phi_i^\dag \, \bar{3}_{\!\scriptscriptstyle L} $ terms
are naively allowed for all the down type quarks. Compare against Eq.(\ref{tykw}), the 
couplings, if allowed, might need some fine-tuning to keep the down and strange quarks light.
In fact, a splitting in mass between $d$ and $D$ as well as $s$ and $S$ is necessary for
fitting SM phenomenology. This is an important issue to be investigated in detail. Similar to
the case of the bottom quark, up and charm quark have their masses from couplings of the 
form $1_{\!\scriptscriptstyle L} \, \Phi_i \Phi_j\, \bar{3}_{\!\scriptscriptstyle L} $ at lowest
order. These structures are not enough to produce the hierarchical quark mass pattern. 
However, the little Higgs structure has considerations of a different global $SU(3)$ for each
$\Phi_i$ (which is not respected by the gauge symmetry itself). One may consider a full
description of these global symmetries for the full Lagrangian and see if they can be used
to help getting a more viable phenomenology without employing un-natural small couplings
for some of the terms here discussed admitted by the gauge symmetry along. As for the
leptonic sector, the representation structure is family universal. The lowest order admissible
coupling is of the form $\ell^+ \,\Phi_i^\dag \Phi_j^\dag \,L$ for $\ell^+$ and $L$ 
representing the singlet and triplet leptons. Neutrino mass constructions looks nontrivial,
and may have to be considered at loop level. The above is limiting the scalar sector to
the two $\Phi_i$'s. One may consider adding extra $SU(3)_L\times U(1)_X$ to 
$SU(2)_L\times U(1)_Y$ symmetry breaking Higgs multiplets, which would likely change
the picture. Care would have to be taken to ensure the little Higgs mechanism is
preserved intact though.  Our discussion on the phenomenology for
the fermions has to stop here in this short letter.

In summary, we have presented a consistent fermionic sector model of
$SU(3)_C\times SU(3)_L\times U(1)_X$ with chiral fermions giving rise to quarks and leptons
of the three-family SM plus some extra quarks and neutrinos that are singlets under
electroweak symmetry. The extra quarks are a top-like quark and two down-sector quarks. 
Group theoretically speaking, the model is a simple twist of the old {331} model with the same
gauge symmetry. The new model is however motivated by solving the hierarchy problem using
the little Higgs mechanism, and seems also to be having more desirable phenomenological 
properties apart from that. It may provide, from the theoretical point of view, a TeV scale 
solution that address both the hierarchy problem and the flavor problem successfully; and may
also gives feasible solution to the experimentally required beyond SM properties of neutrinos 
such as mass oscillation. In our opinion, more model-building of the type, and careful 
phenomenological studies of the successful models, deserve up most attention from particle
physics. 

{\it NOTE:} After posting the first draft of our manuscript, we came to realize that the {331}
model fermion spectrum obtained here has actually been available in the literature\cite{SVS}.
These earlier works having no connection with the little Higgs idea which motivates our
rediscovery though. Accordingly, the scalar spectrum and Yukawa couplings discussed 
are not the same as that discussed here, as required by the little Higgs mechanism.

Our work is partially supported by the National Science Council of Taiwan, under
grant number NSC91-2112-M-008-042.

\end{document}